\documentclass{elsart}

\usepackage{natbib}

\usepackage{amssymb}

\usepackage{graphicx}

\begin{document}

\begin{frontmatter}

\title{Stochastic Resonance in an Extended FitzHugh-Nagumo System:
the Role of Selective Coupling}

\author[label1]{Claudio J. Tessone,}
\ead{tessonec@imedea.uib.es}
\author[label2]{Horacio S. Wio\corauthref{cor1}}
\ead{wio@ifca.unican.es} \corauth[cor1]{Corresponding Author: H.S.
Wio, Instituto de F\'isica de Cantabria, 39005 Santander, Spain}
\address[label1]{Institut Mediterrani d'Estudis Avan\c{c}ats
(IMEDEA-CSIC), Universitat de les Illes Balears, E07122 Palma de
Mallorca, Spain.}
\address[label2]{Instituto de F\'{\i}sica de Cantabria, Universidad
de Cantabria and CSIC, E39005 Santander, Spain.}

\begin{abstract}
Here we present a study of \textit{stochastic resonance} in an
extended FitzHugh-Nagumo system with a field dependent activator
diffusion. We show that the system response (here measured through
the output signal-to-noise ratio) is enhanced due to the particular
form of the non-homogeneous coupling. Such a result supports
previous ones obtained in a simpler scalar reaction-diffusion system
and shows that such an enhancement, induced by the field dependent
diffusion -or selective coupling-, is a robust phenomenon.
\end{abstract}


\end{frontmatter}

\section{Introduction}

\textit{Stochastic resonance} (SR) is one of the most interesting
\textit{noise-induced phenomena} that arises from the interplay
between \textit{deterministic} and \textit{random} dynamics in a
\textit{nonlinear} system \cite{RMP}. A large number of examples
showing SR occur in \textit{extended} systems: for example, diverse
experiments were carried out to explore the role of SR in sensory
and other biological functions \cite{biol} or in chemical systems
\cite{sch}. These, together with the possible technological
applications, motivated many recent studies showing the possibility
of achieving an enhancement of the system response by means of the
coupling of several units in what conforms an \textit{extended
medium} \cite{extend1,extend2,extend3}.

In previous works \cite{extend2,extend3} we have studied the
stochastic resonant phenomenon in extended systems, when transitions
between two different spatial patterns occurs, exploiting the
concept of the \textit{non-equilibrium potential} (NEP)
\cite{GR,I0}: a Lyapunov functional of the associated deterministic
system that, for non-equilibrium systems, plays a role similar to
that of a thermodynamic potential in equilibrium thermodynamics.
Such NEP characterizes the global properties of the dynamics:
attractors, relative (or nonlinear) stability of these attractors,
height of the barriers separating attraction basins and, in
addition, allowing us to evaluate the transition rates among the
different attractors. In another work \cite{extend3c} we have also
shown that, for a scalar reaction-diffusion system with a
density-dependent diffusion and a known form of the NEP, the
non-homogeneous spatial coupling changes the effective dynamics of
the system and contributes to enhance the SR phenomenon.

Here we report on a study of SR in an extended system: an array of
FitzHugh-Nagumo \cite{FHN} units, with a density-dependent
(diffusive-like) coupling. The NEP for this system was found within
the excitable regime and for particular values of the coupling
strength \cite{extend3}. In the general case, however, the form of
the NEP has not been found yet. Nevertheless, the idea of the
existence of such a NEP is always \textit{underlying} our study.
Hence, we have resorted to an study based on numerical simulations,
analyzing the influence of different parameters on the system
response. The results show that the enhancement of the
signal-to-noise ratio found for a scalar system \cite{extend3c} is
robust, and that the indicated non-homogeneous coupling could
clearly contribute to enhance the SR phenomenon in more general
situations.

\section{Theoretical Framework}

\subsection{The Model}

For the sake of concreteness, we consider a simplified version of
the FitzHugh-Nagumo \cite{extend3,I0,FHN} model. This model has been
useful for gaining qualitative insight into the excitable and
oscillatory dynamics in neural and chemical systems \cite{LGN04}. It
consist of two variables, in one hand $u$, a (fast) activator field
that in the case of neural systems represents the voltage variable,
while in chemical systems represents a concentration of a
self-catalytic species. On the other hand $v$, the inhibitor field,
associated with the concentration of potassium ions in the medium
(within a neural context), that inhibits the generation of the $u$
species (in a chemical reaction). Instead of considering the usual
cubic like nonlinear form, we use a piece-wise linear version
\begin{eqnarray}
\label{eq:ucont}\epsilon \, \frac{\partial u(x,t)}{\partial t} &=&
\frac{\partial }{\partial x} \left( D_u(u) \, \frac{\partial
u}{\partial x} \right) + f(u) - v + \xi(x,t) \\
\frac{\partial v(x,t)}{\partial t} &=&
\label{eq:vcont}\frac{\partial }{\partial x} \left( D_v(v) \,
\frac{\partial u}{\partial x}\right) +  \beta\,u - \alpha\, v,
\end{eqnarray}
where $f(u) = -u + \Theta(u-\phi_c) $, and $\xi(x,t)$ is a
$\delta$-correlated white Gaussian noise, that is $\left< \xi(x,t)
\right> = 0$ and $ \left< \xi(x,t) \xi(x',t') \right> = 2 \gamma
\delta(x-x') \delta(t-t')$. Here $\gamma$ indicates the noise
intensity and $\phi_c $ is the ``discontinuity" point, at which the
piece-wise linearized function $f(u)$ presents a jump. In what
follows, the parameters $\alpha $ and $\beta $ are fixed as $\alpha
=0.3$ and $\beta = 0.4$. Finally, $\epsilon$ is the parameter that
indicates the time-scale ratio between activator and inhibitor
variables, and is set as $\epsilon=0.03$. We consider Dirichlet
boundary conditions at $x=\pm L$. Although the results are
qualitatively the same as those that could appear considering the
usual FitzHugh-Nagumo equations, this simplified version allows us
to compare directly with the previous analytical results for this
system \cite{extend3}.

\begin{figure} 
\centering \label{fig:pattern}
\includegraphics[width=7cm,angle=-90]{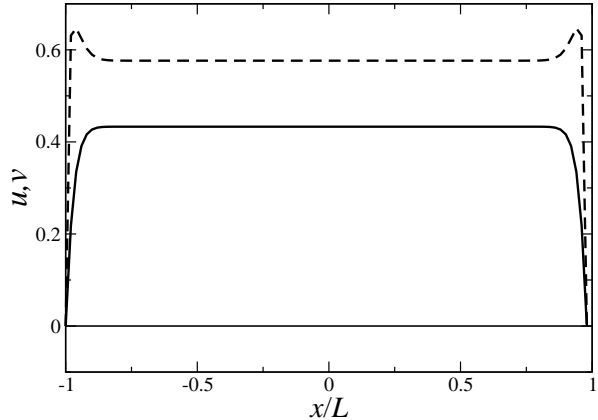}
\caption{ We show the stable patterns that arise in the system.
There is one stable pattern that is identically zero, i.e.
$P^u_0(x)=P^v_0(x)=0$ and another which is non-zero ($P^u_1(x)$,
$P^v_1(x)$). The patterns for the fields $u(x)$ and $v(x)$ are
plotted in dashed and solid lines, respectively. The parameters, are
$D_u= 0.3$, $D_v= 1$, $h=2$.}
\end{figure}

As in \cite{extend3c}, we assume that the diffusion coefficient
$D_u(u)$ is not constant, but depends on the field $u$ according to
$D_u(u) = \, D_u \left[ 1 + h \, \Theta(u-\phi_c) \right] $. This
form implies that the value of $D_u(u)$ depends ``selectively" on
whether the field $u$ fulfills $u > \phi_c$ or $u < \phi_c$. $D_u$
is the value of the diffusion constant without such ``selective"
term, and $h$ indicates the size of the difference between the
diffusion constants in both regions (clearly, if $h=0$ then
$D_u(u)=D_u$ constant). $D_v(v)$ is the diffusion for the inhibitor
$v$, that here we assume to be homogeneously constant.

It is worthwhile noting that when the parameter $h$ is large enough,
under some circumstances the coupling term might become negative.
This is what is known as ``inhibitory coupling" \cite{Dayan}. This
is a very interesting kind of coupling that has attracted much
attention in the last years, both in neural and chemical context,
that we will not discuss here.

This system is known to exhibit two stable stationary patterns. One
of them is $u(x)=0$, $v(x)=0$, while the other is one with non-zero
values and can be seen in Fig. 1. We will denote with $P^{u,v}_0(x)$
and $P^{u,v}_1(x)$, the patterns for $u$ and $v$ fields. Further, we
consider that an external, periodic, signal enters into the system
through the value of the threshold $\phi_c$,
\begin{equation}
\phi_c(t) = \phi_c + \delta\phi \, \cos (\omega t),
\end{equation}
where $\omega$ is the signal frequency, and $\delta\phi $ its
intensity.

All the results shown in this paper were obtained through numerical
simulations of the system. The second order spatially discrete
version of the system indicated in Eqs. (\ref{eq:ucont},
\ref{eq:vcont}) reads
\begin{eqnarray}
\dot{u_i} &=& D_{u,i} ( u_{i-1} + u_{i+1}-2 u_{i} ) +
(D_{u,i+1}-D_{u,i-1}) ( u_{i+1} + u_{i-1} ) \nonumber \\
& & \label{eq:disc1} \hspace{8cm}+ f(u_i) - v_i + \xi _{i}(t) \\
\label{eq:disc2}\dot{v_i} &=& D_{v} ( v_{i-1} + v_{i+1}-2 v_{i} ) +
\beta\, u_i - \alpha\,v_i.
\end{eqnarray}
We have performed extensive numerical simulations of this set of
equations exploiting the Heun's algorithm \cite{1999_nises}.

\subsection{Response's Measures}

Since the discovery of the stochastic resonance phenomenon, several
different forms of characterizing it have been introduced in the
literature. Some examples are: (i) output signal-to-noise ratio
(SNR) \cite{RMP,McNM}, (ii) the spectral amplification factor (SAF)
\cite{jung1,jung2}, (iii) the residence time distribution
\cite{haenggi1,gamma1}, and, more recently, (iv) information theory
based tools \cite{bulsladri,neiman,noso}. Along this paper, we will
use the output SNR at the driving frequency $\omega$.

In this spatially-extended system, there are different ways of
measuring the overall system response to the external signal. In
particular, we evaluated the output SNR in three different ways (the
units being given in dB)
\begin{itemize}
\item SNR for the element $N/4$ of the chain evaluated over the
dynamical evolution of $u_{N/4}$, that we call $\textit{SNR}_1$.
\item SNR for the middle element of the chain evaluated over the
dynamical evolution of $u_{N/2}$, that we call $\textit{SNR}_2$.
Having Dirichlet boundary conditions, the local response depends on
the distance to the boundaries.
\item In order to measure the overall response of the system to
the external signal, we computed the SNR as follows: We digitized
the system dynamics to a dichotomic process $s(t)$: At time $t$ the
system has an associated value of $s(t)=1\, (0)$  if the Hilbert
distance to pattern $1\, (0)$ is lower than to the other pattern.
Stated in mathematical terms, we computed the distance
$\mathcal{D}_2[\cdot,\cdot]$ defined by
$$\mathcal{D}_2[f,g]=\left( \int_{-L}^L dx\, \left( f(x)-g(x)
\right)^2 \right)^{1/2}$$ in the Hilbert space of the real-valued
functions in the interval $[-L,L]$, i.e. $\mathcal{L}_2$. At time
$t$, a digitized process is computed by means of
\begin{equation}
s (t) = \left\{ %
\begin{array}{c}
1 \qquad \hbox{if } \mathcal{D}_2
\left[P^u_1(x),u(x,t)\right]<\mathcal{D}_2 \left[
P^u_0(x),u(x,t)\right]\\ 0 \qquad \hbox{if } \mathcal{D}_2
\left[P^u_1(x),u(x,t)\right]\geq\mathcal{D}_2 \left[ P^u_0(x),u(x,t)
\right]
\end{array}  \right. ,
\end{equation}
We call this measure $SNR_{p}$.
\end{itemize}

\section{Results}

As indicated above, Eqs. (\ref{eq:disc1}) and (\ref{eq:disc2}) have
been integrated by means of the Heun method \cite{TSM}. We have
fixed the parameters $\epsilon=0.03$, $\phi_c=0.52$ and adopted an
integration step of $\Delta t=10^{-3}$. For the signal frequency we
adopted $\omega=2 \pi / 3.2 = 1.9634295\ldots$. The simulation was
repeated 250 times for each parameter set, and the SNR was computed
by recourse of the average power spectral density.

\begin{figure} 
\centering
\includegraphics[width=4.5cm,angle=-90]{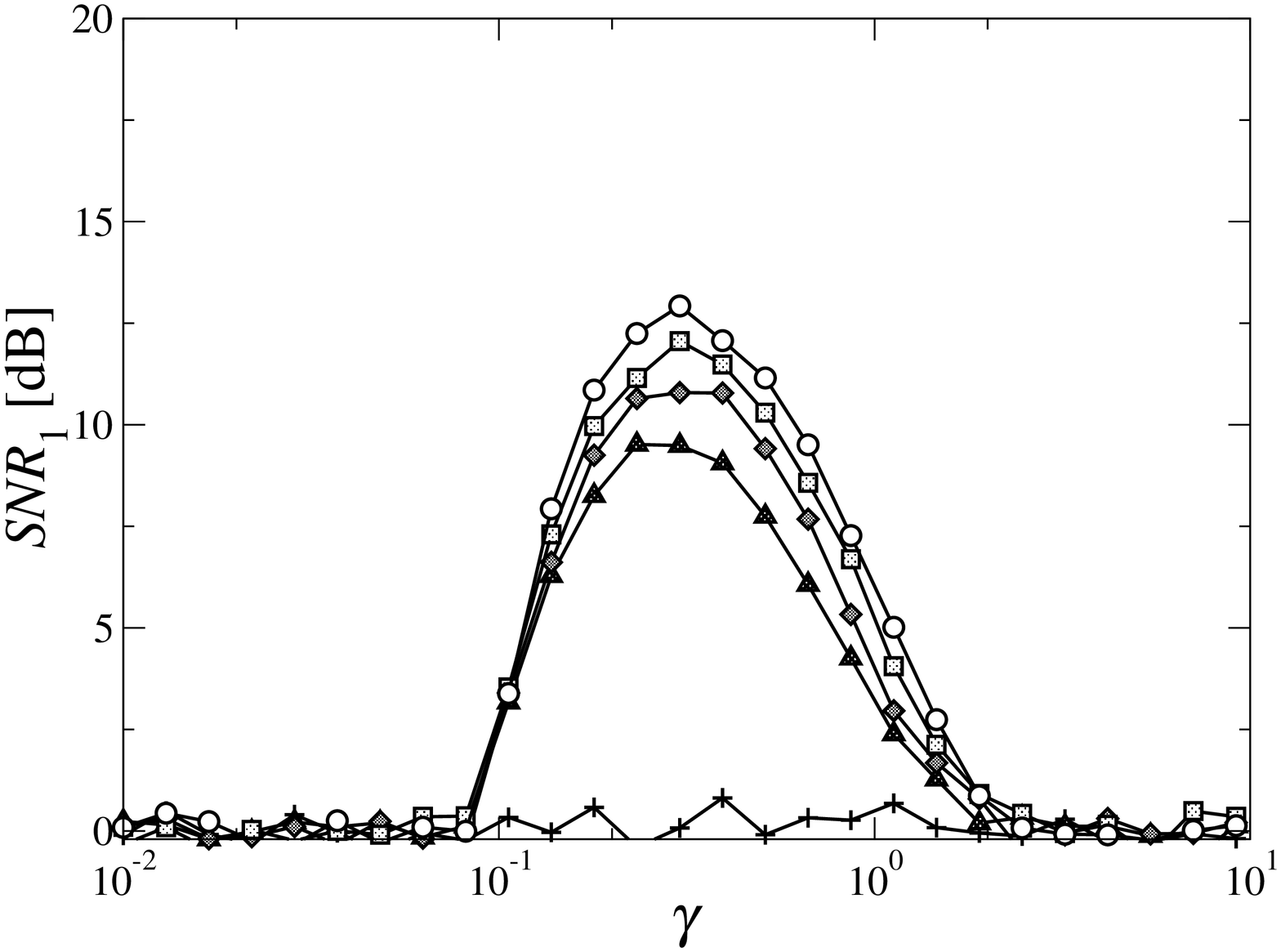}
\includegraphics[width=4.5cm,angle=-90]{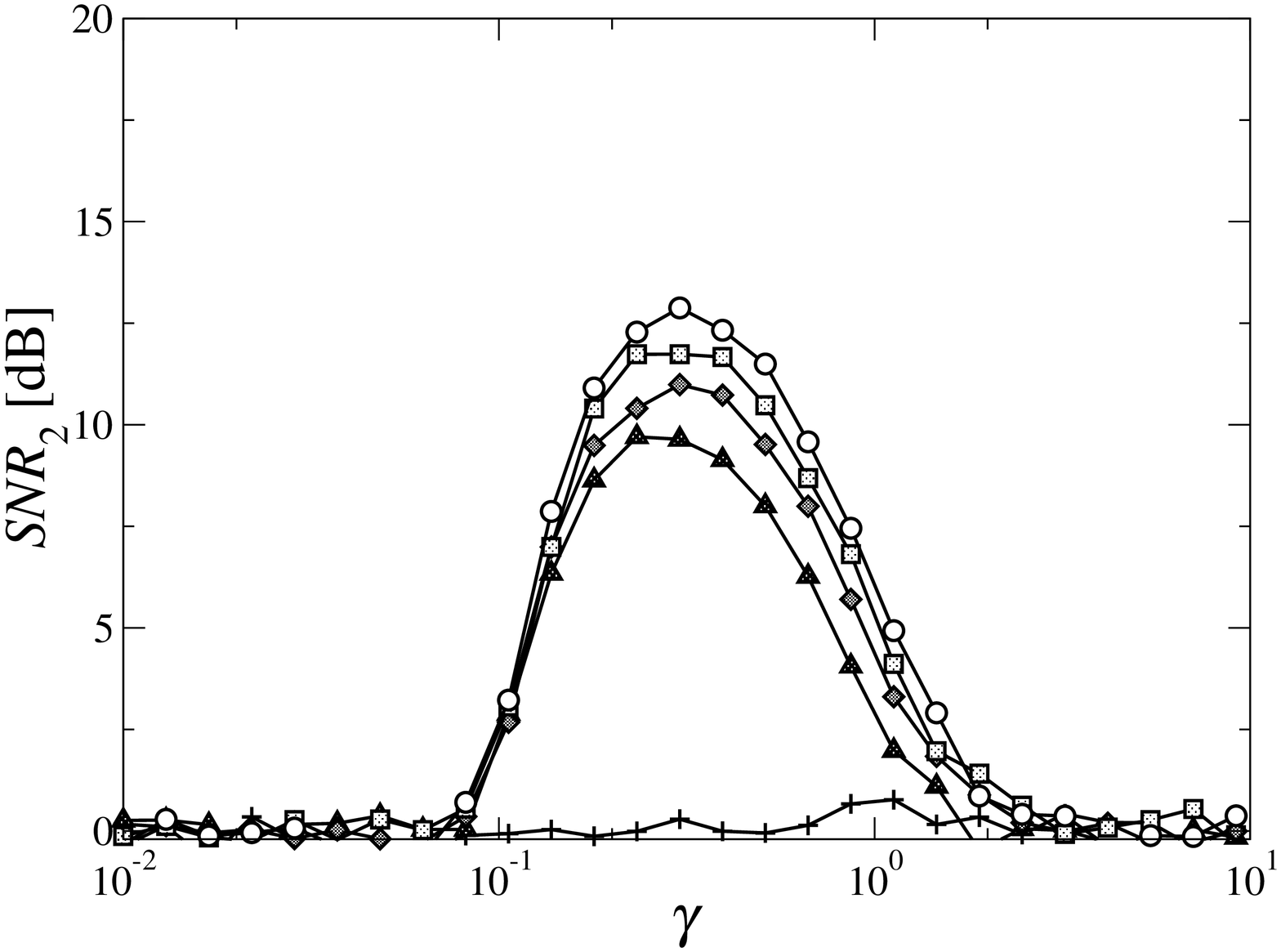}
\includegraphics[width=4.5cm,angle=-90]{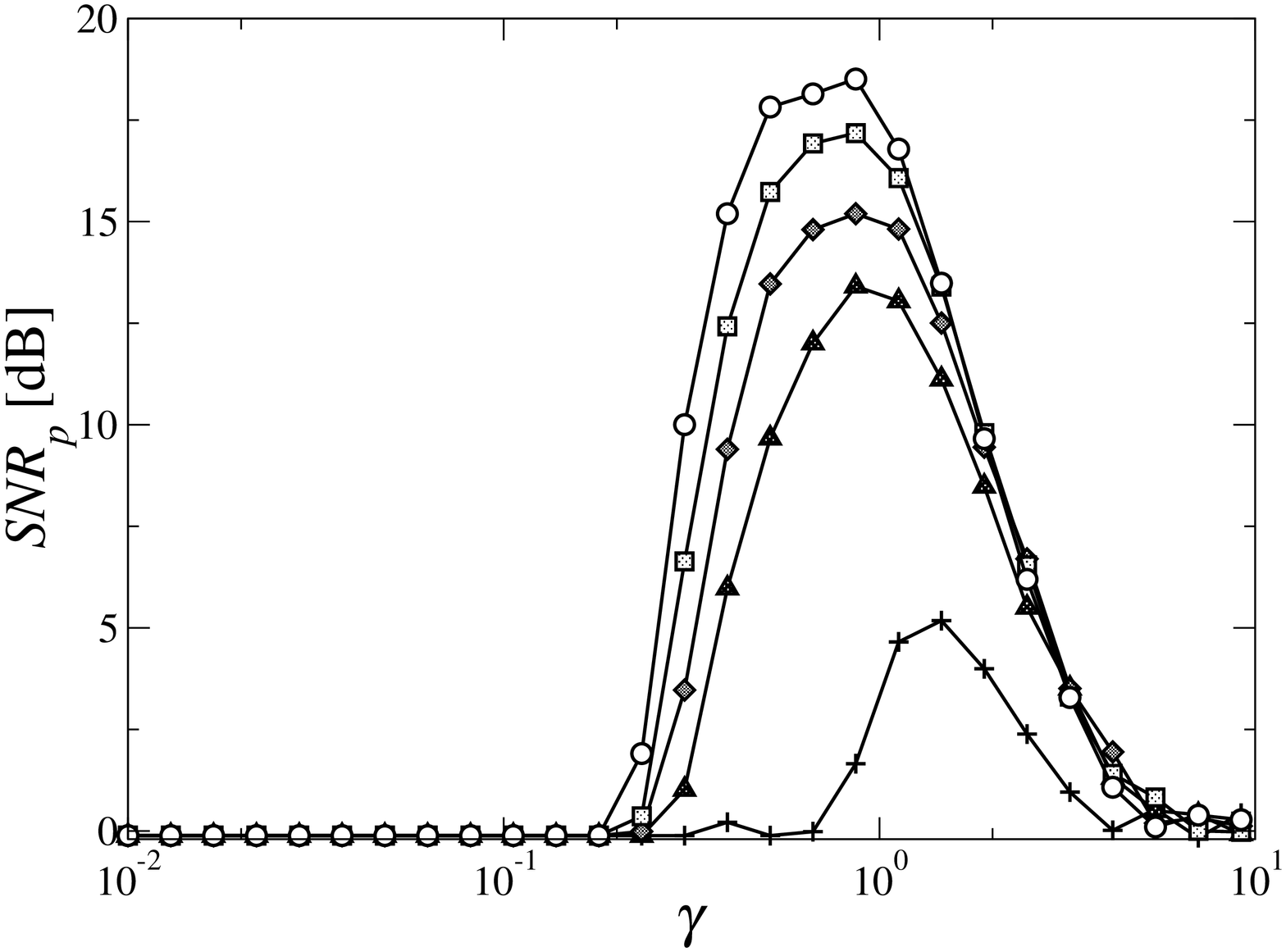}
\caption{ SNR vs. $\gamma$, the noise intensity, for the three
different measures we use. The parameters are $\delta\phi= 0.4$,
$D_v= 1.$, $\omega=2\pi/3.2$, $N=51$. The different curves represent
different values of $h$, showing an enhancement of the response to
the external signal for $\gamma>0$. In particular it is shown:
$h=-2$ ($+$), $h=-1$ ($\triangle$), $h=0$ ($\diamond$), $h=1$
($\square$) and $h=2$ ($\bigcirc$). }\label{SNR-gamma}
\end{figure}

Figure \ref{SNR-gamma} depicts the results for the different SNR's
measures we have previously defined as function of the noise
intensity $\gamma$. We adopted the following values: $\delta \phi =
0.4$, $D_v = 1.$ and $N = 51$. In all three cases it is apparent
that there is an enhancement of the response for $h > 0$, when
compared with the $h = 0$ case, while for $h < 0$ the response is
smaller.

\begin{figure} 
\centering
\includegraphics[width=4.5cm,angle=-90]{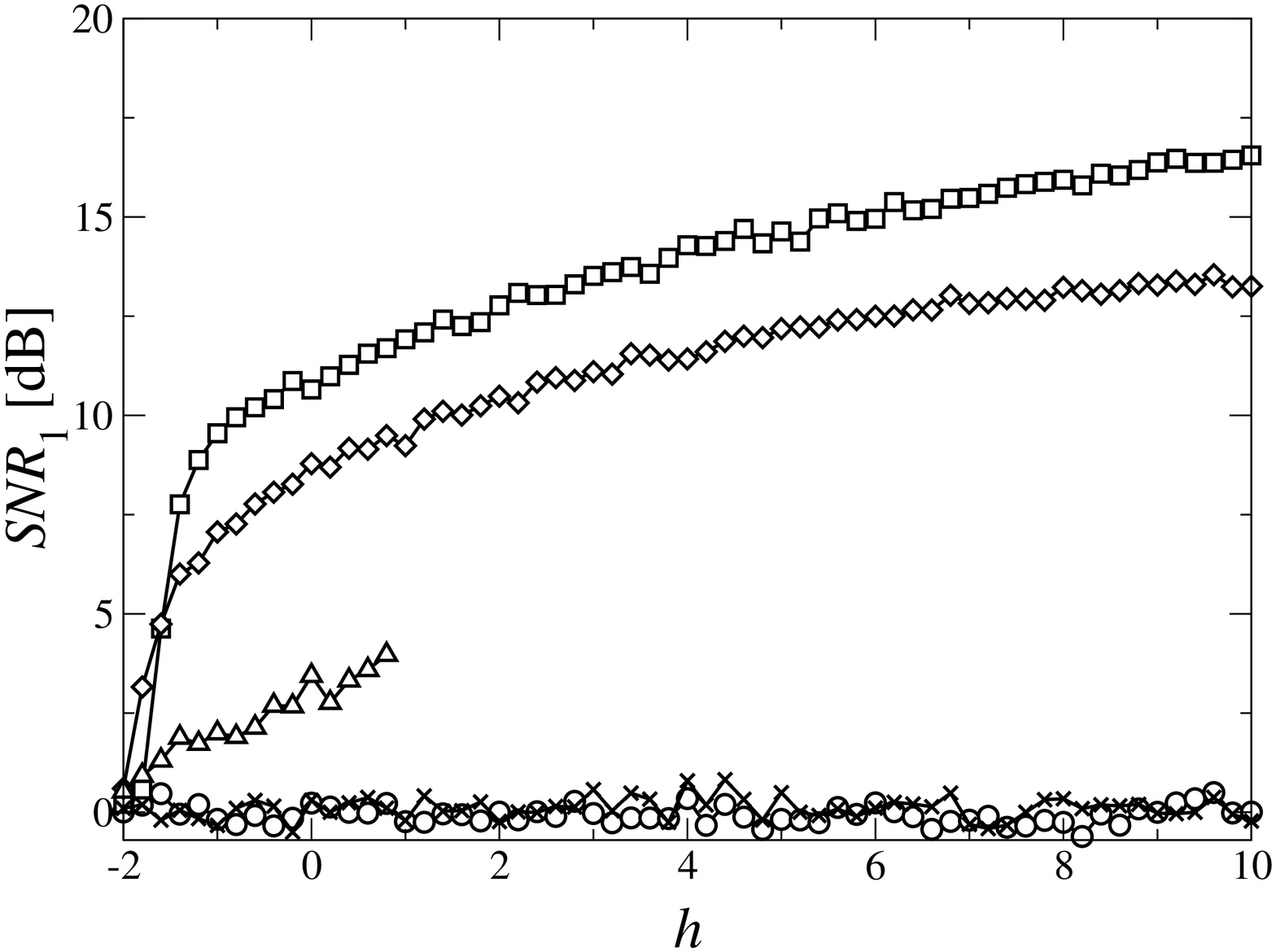}
\includegraphics[width=4.5cm,angle=-90]{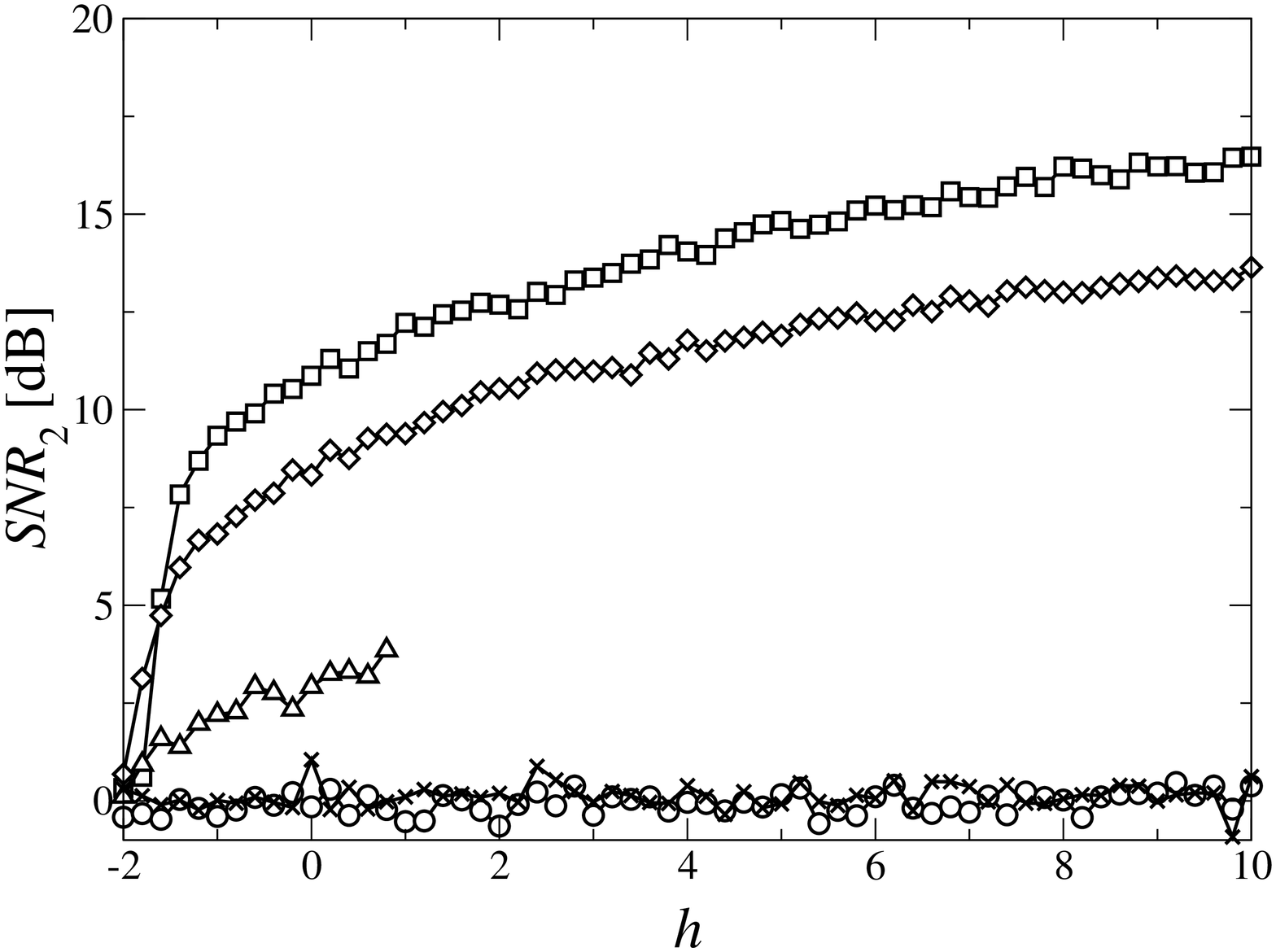}
\includegraphics[width=4.5cm,angle=-90]{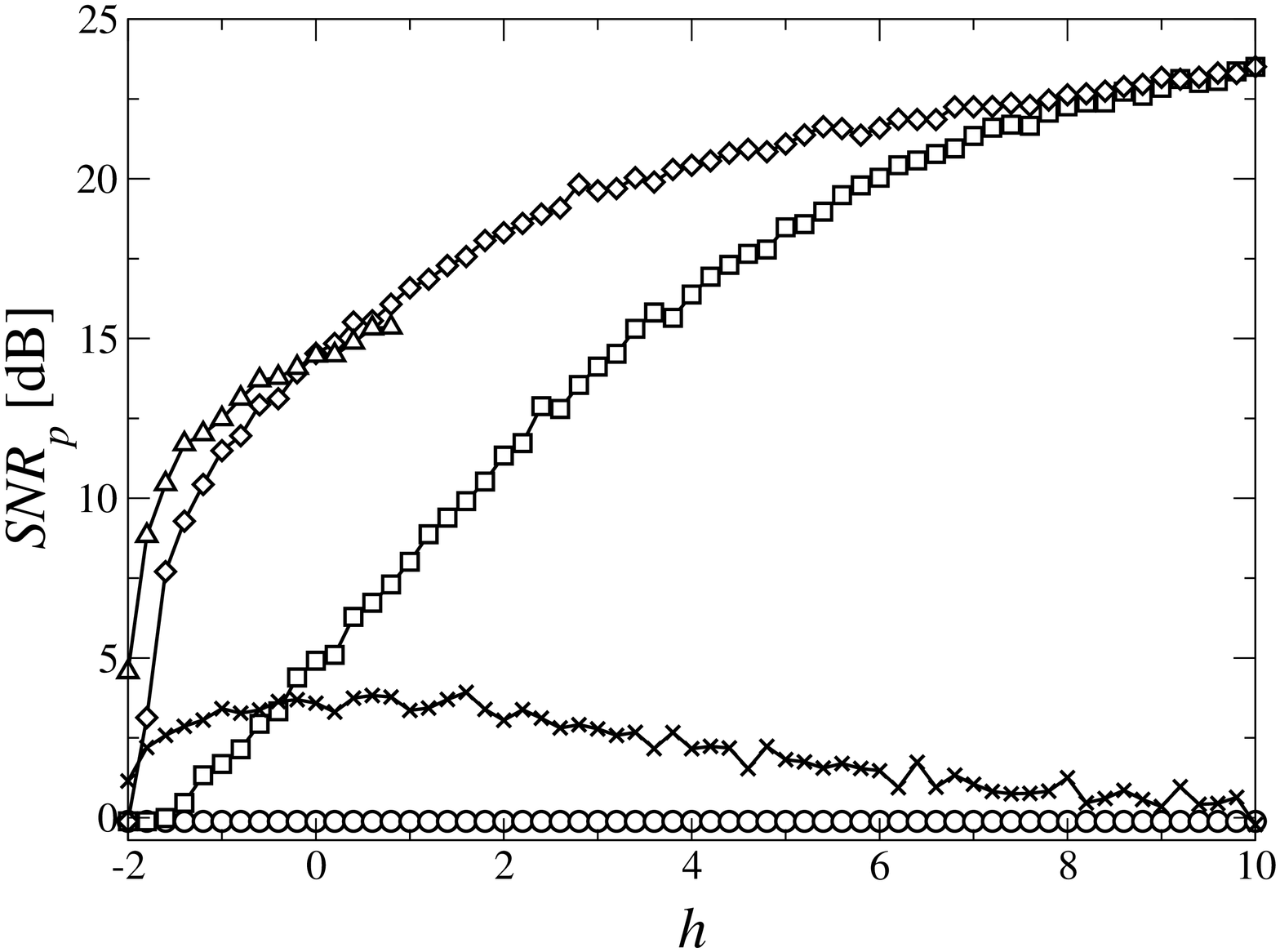}
\caption{SNR vs. $h$, the selectiveness of coupling, for the three
different measures we use. The parameters are $\delta\phi= 0.4$,
$D_v= 1.$, $\gamma = 0.032~(\bigcirc), 0.32~(\square),
0.6~(\diamond), 1.2~(\square), 3.2~(\times)$, $D_u=0.3$,
$N=51$.}\label{SNR-h}
\end{figure}

In Fig. \ref{SNR-h} we show the same three response's measures, but
now as a function of $h$. We have plotted the maximum of each SNR
curve, for three different values of the noise intensity, and for
$\delta\phi= 0.4$, $D_v= 1.$, $\gamma = 0.01, 0.1, 0.3$, $D_u=0.3$,
and $N=51$. It is clear that there exists an optimal value of
$\gamma$ such that, for such a value, the phenomenon is stronger
(that is, the response is larger). It is apparent the rapid fall in
the response for $h<0$.

\begin{figure} 
\centering
\includegraphics[width=4.5cm,angle=-90]{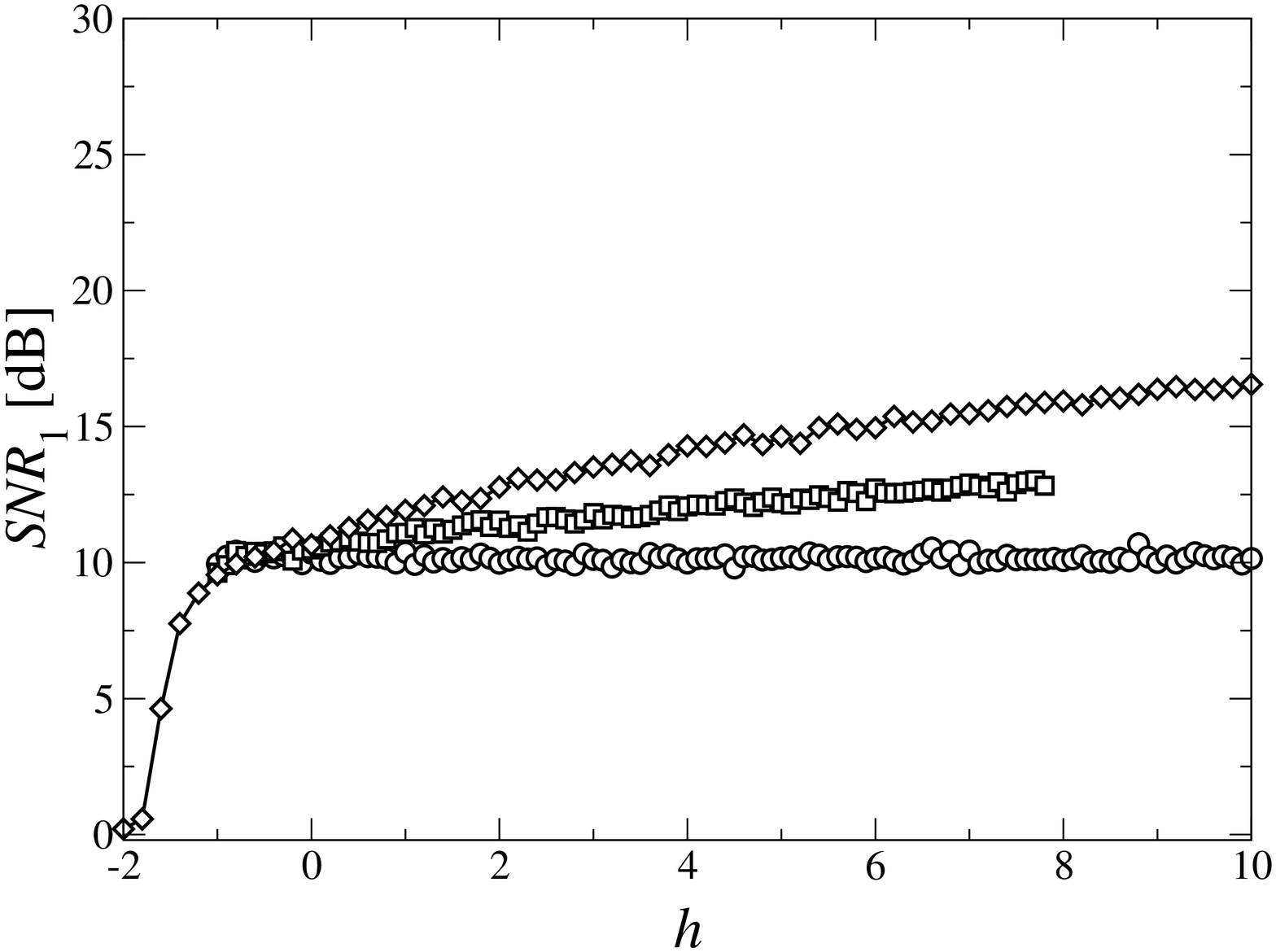}
\includegraphics[width=4.5cm,angle=-90]{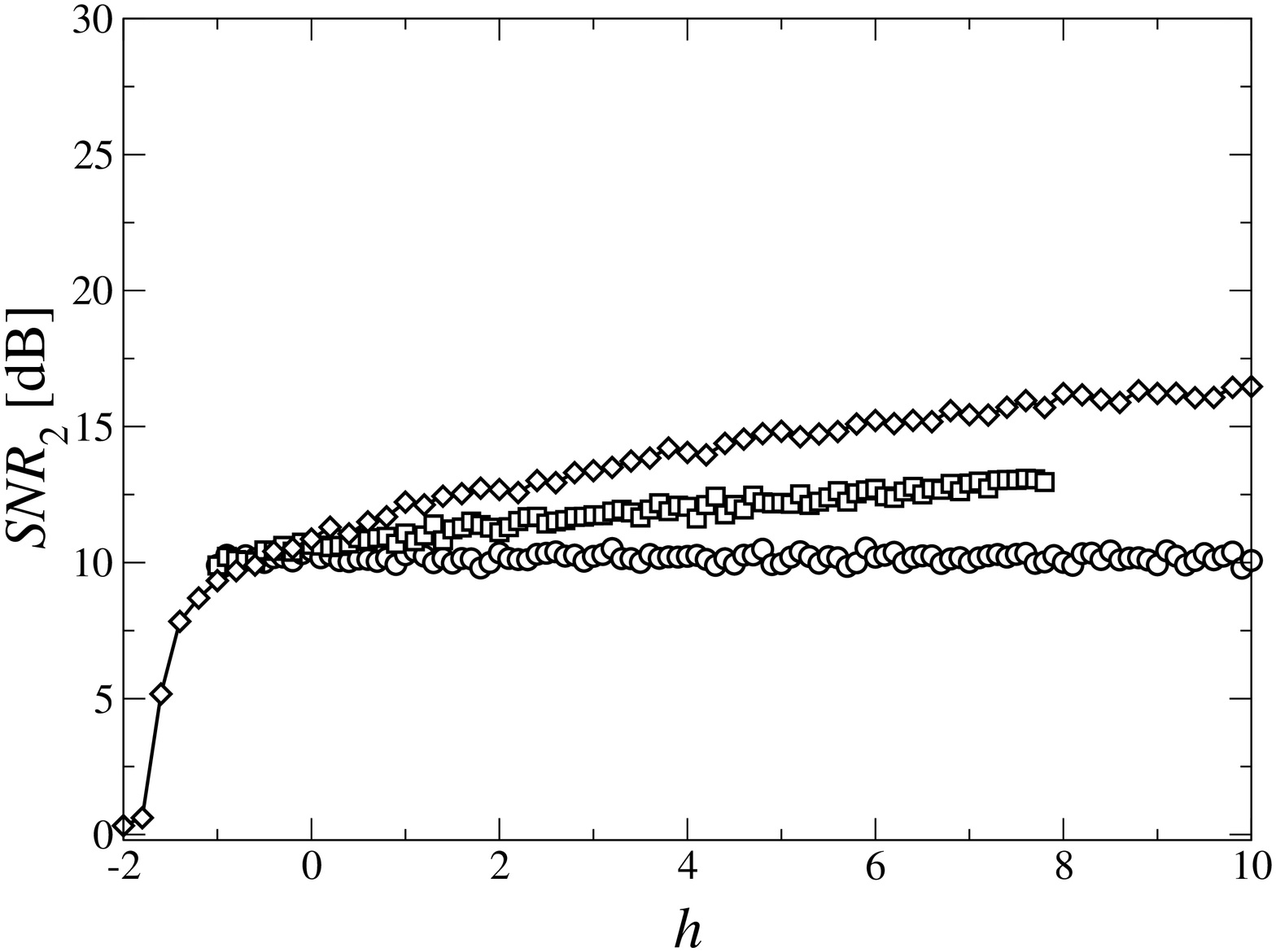}
\includegraphics[width=4.5cm,angle=-90]{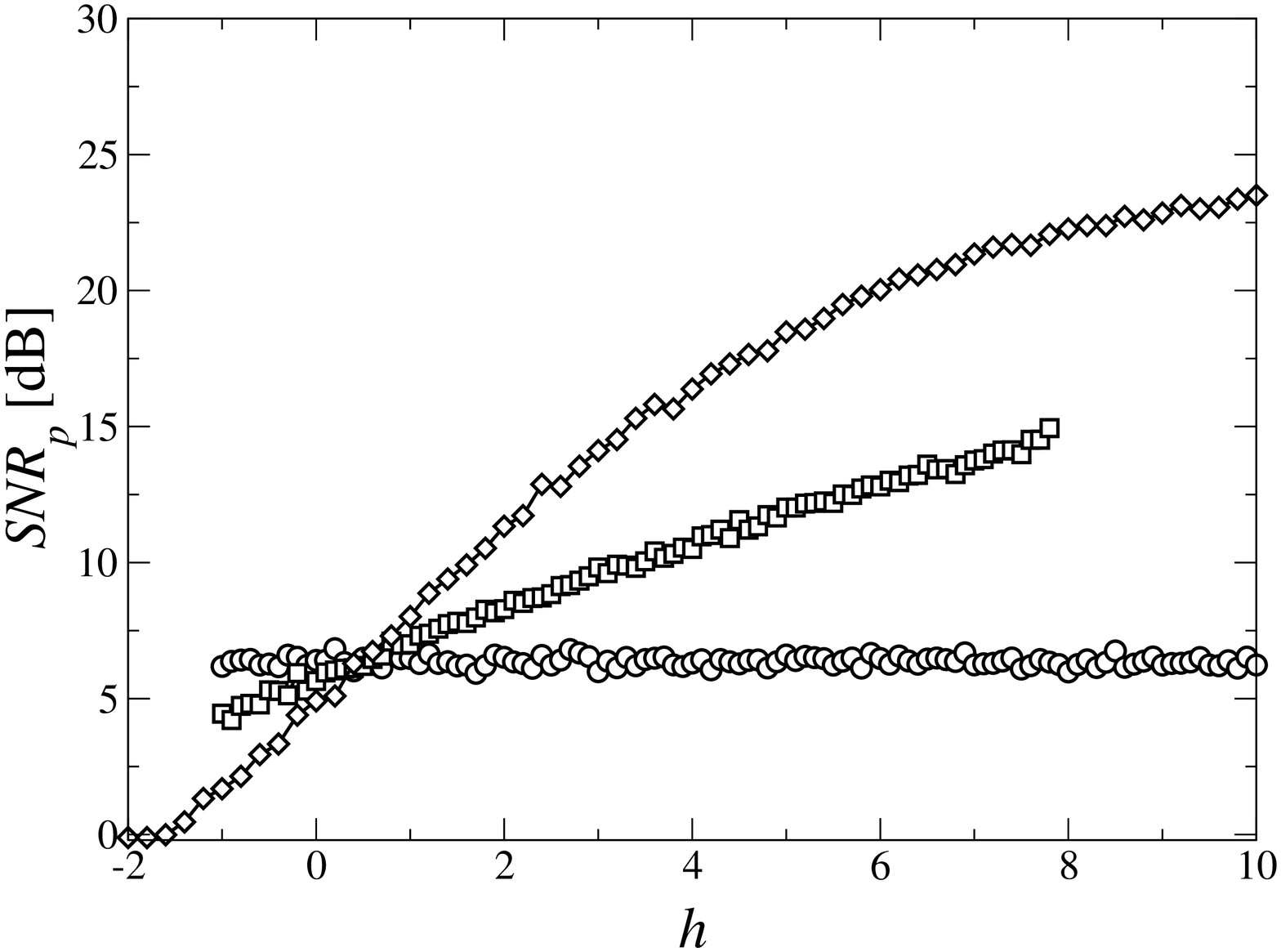}
\caption{SNR vs. $h$, the selectiveness of coupling, for different
values of $D_u$. The parameters are $\delta\phi= 0.4$, $D_v= 1.$,
$\gamma = 0.32$, $D_u=0.0~(\bigcirc), 0.1~(\square),
0.3~(\diamond)$, $N=51$.}\label{SNR-h_dX}
\end{figure}

In figure \ref{SNR-h_dX} we show the dependance of SNR on $h$, for
different values of the diffusion which depends on the activator
density $D_u$. It is apparent that the response becomes larger when
the value of $D_u$ is larger. However, as was discussed in
\cite{extend2,extend3}, it is clear that for still larger values of
$D_u$, the symmetry of the underlying potential (that is the
relative stability between the attractors) is broken and the
response finally falls-down.

\begin{figure} 
\centering
\includegraphics[width=4.5cm,angle=-90]{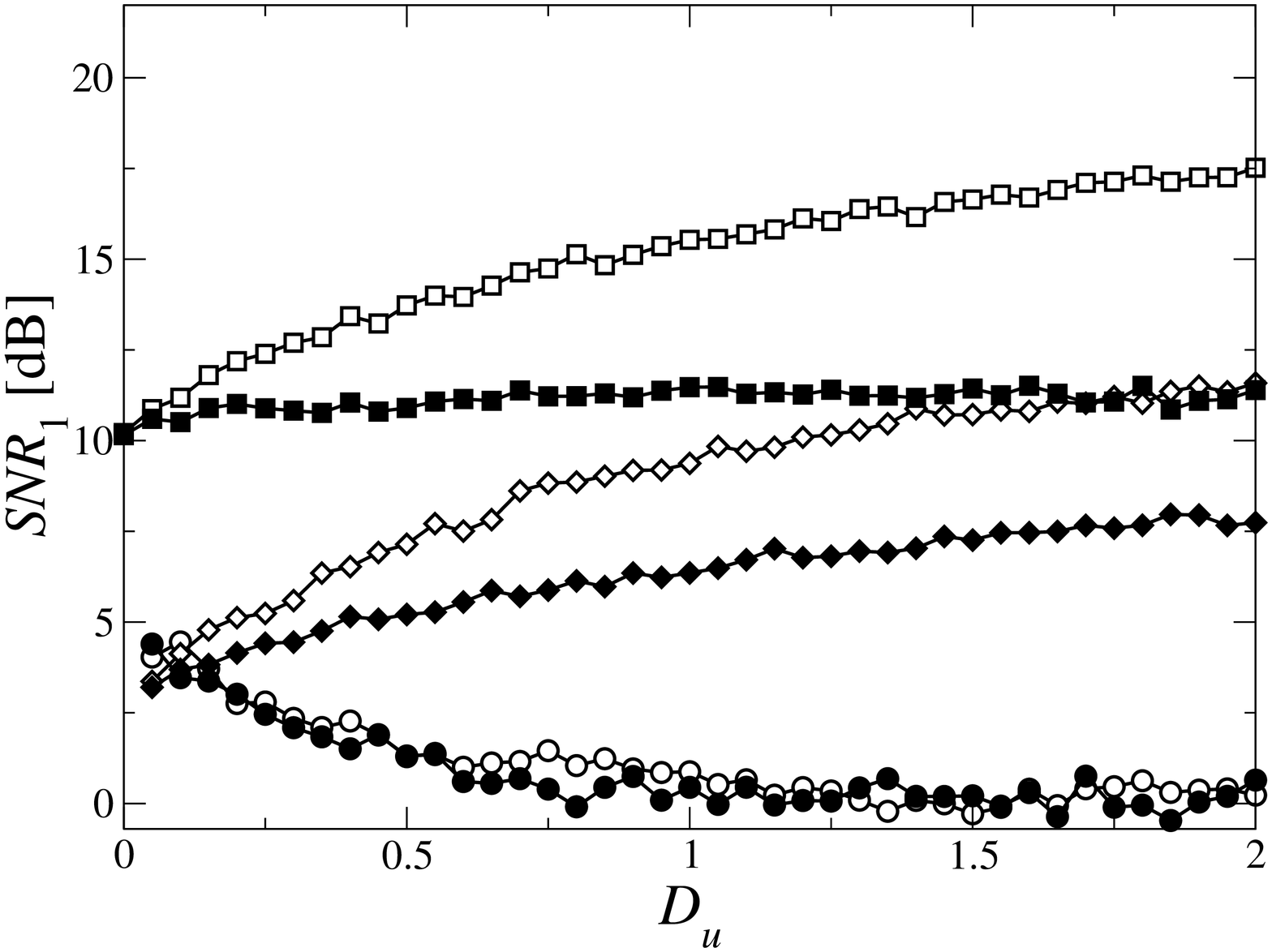}
\includegraphics[width=4.5cm,angle=-90]{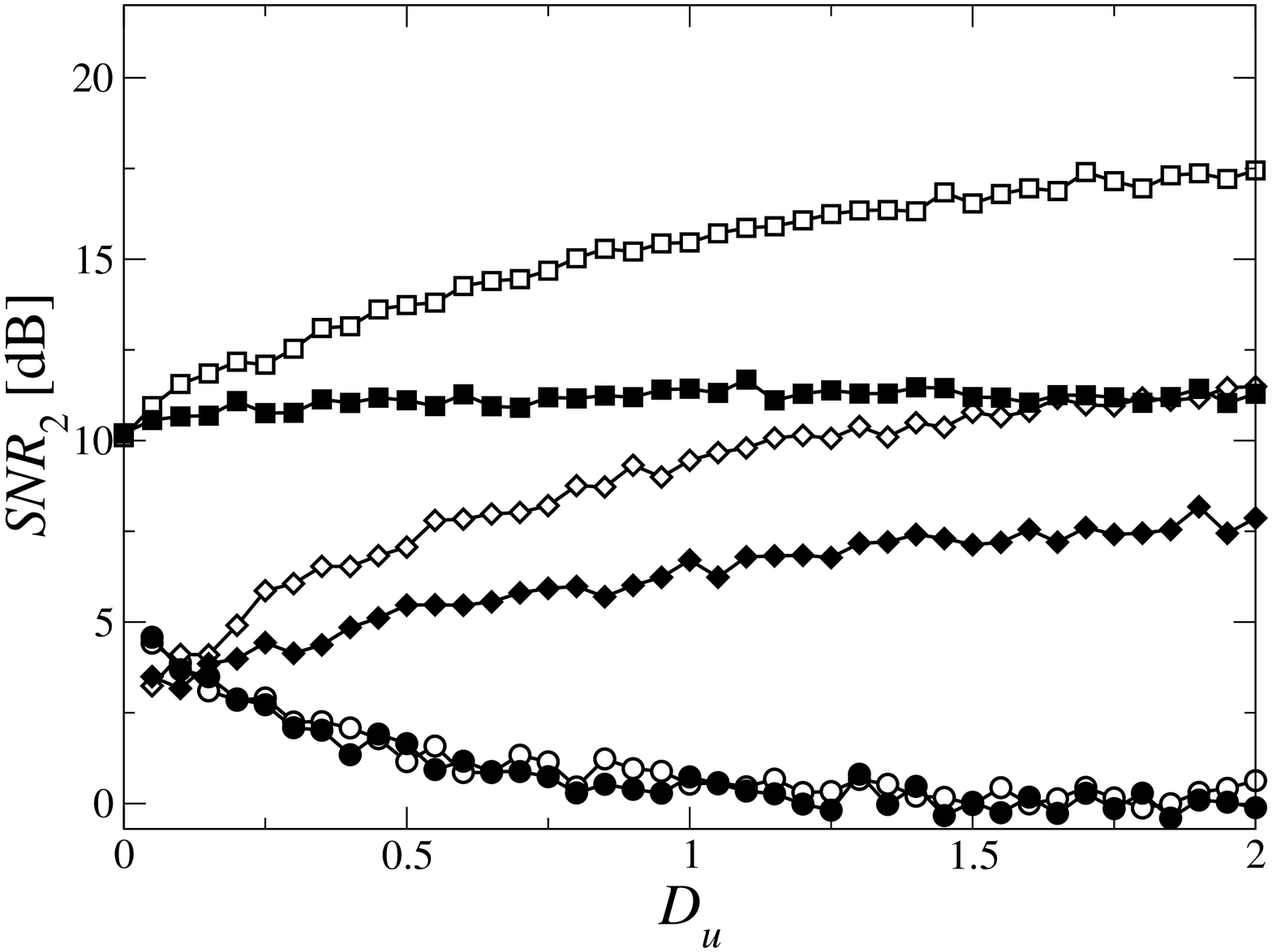}
\includegraphics[width=4.5cm,angle=-90]{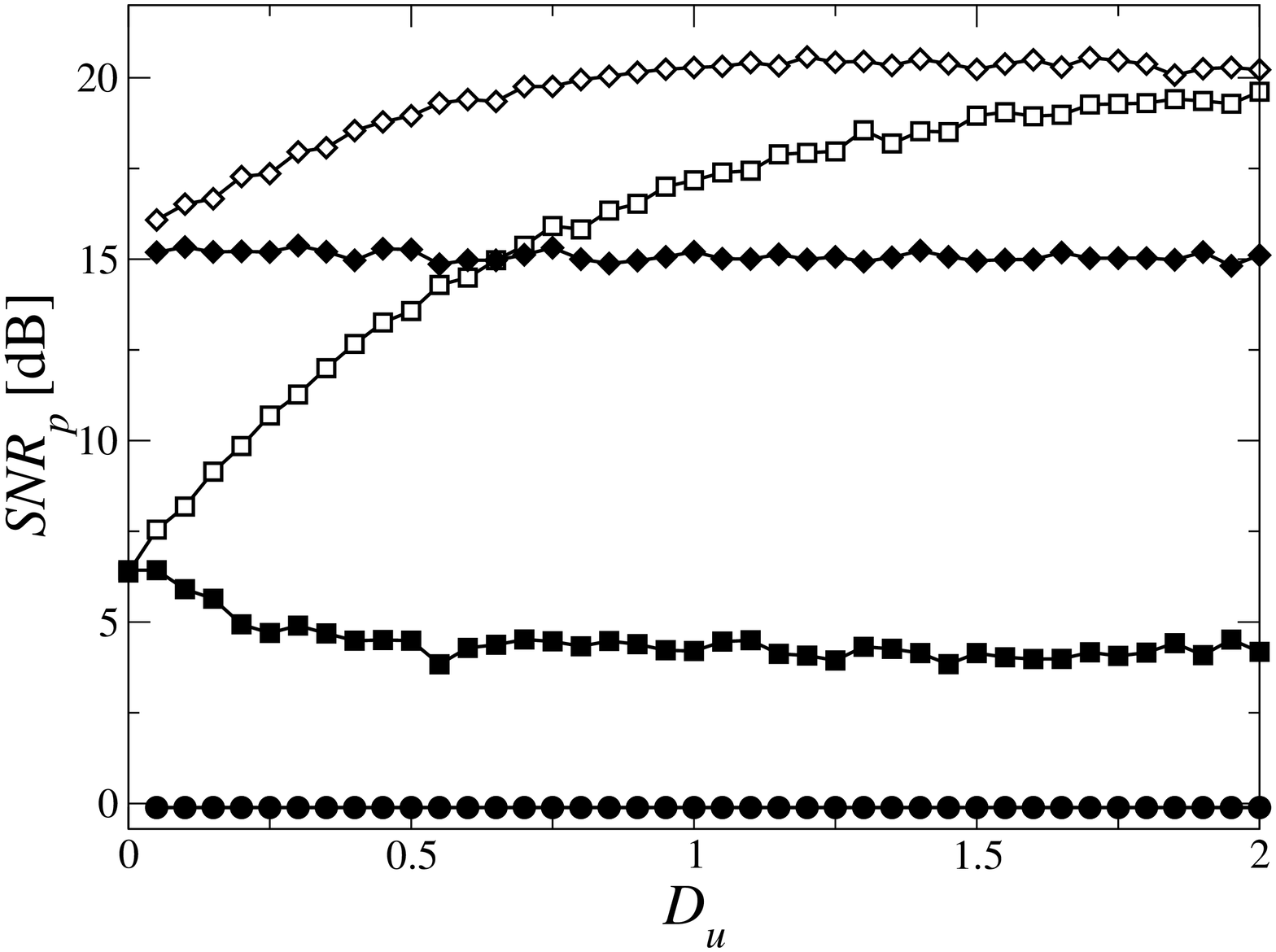}
\caption{SNR vs. $D_u$, the diffusiveness in activator variable $u$,
for the three different measures we use. The parameters are
$\delta\phi= 0.4$, $D_v= 1.$, $\gamma = 0.1~(\bigcirc),
0.32~(\square), 1.0~(\diamond)$, while the white symbols represent
$h=2$ and the black ones, $h=0$. The system size is $N=51$.
}\label{SNR-Du}
\end{figure}

Figure \ref{SNR-Du} shows the results of the SNR, but now as
function of $D_u$, the activator diffusivity, for different values
of $\gamma$, and for $\delta\phi= 0.4$, $D_v= 1.$ and $N=51$. It can
be seen that, independently from the coupling strength $D_u$, the
response to the external signal grows with the selectiveness of the
coupling, showing the robustness of the phenomenon.

\begin{figure} 
\centering
\includegraphics[width=4.5cm,angle=-90]{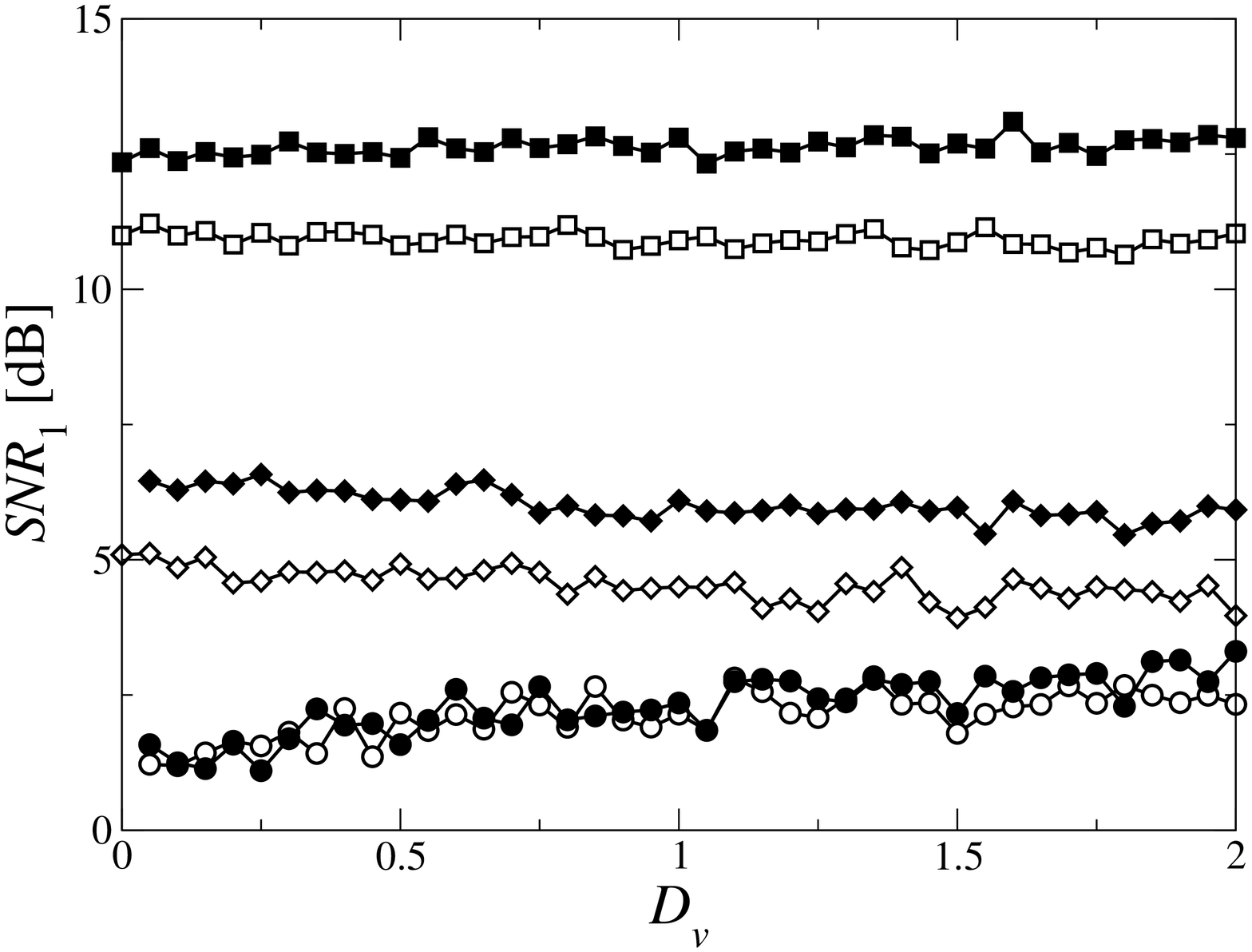}
\includegraphics[width=4.5cm,angle=-90]{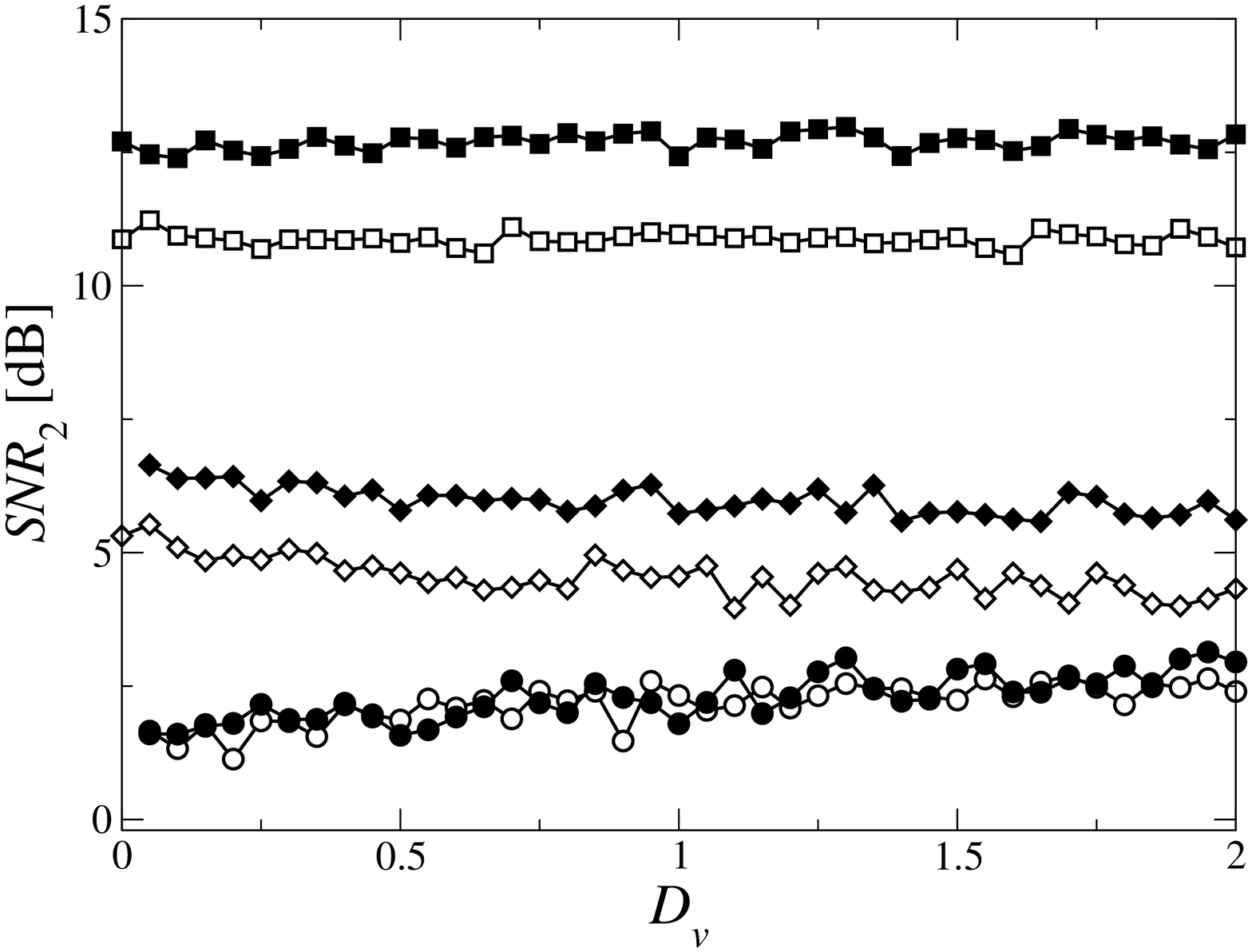}
\includegraphics[width=4.5cm,angle=-90]{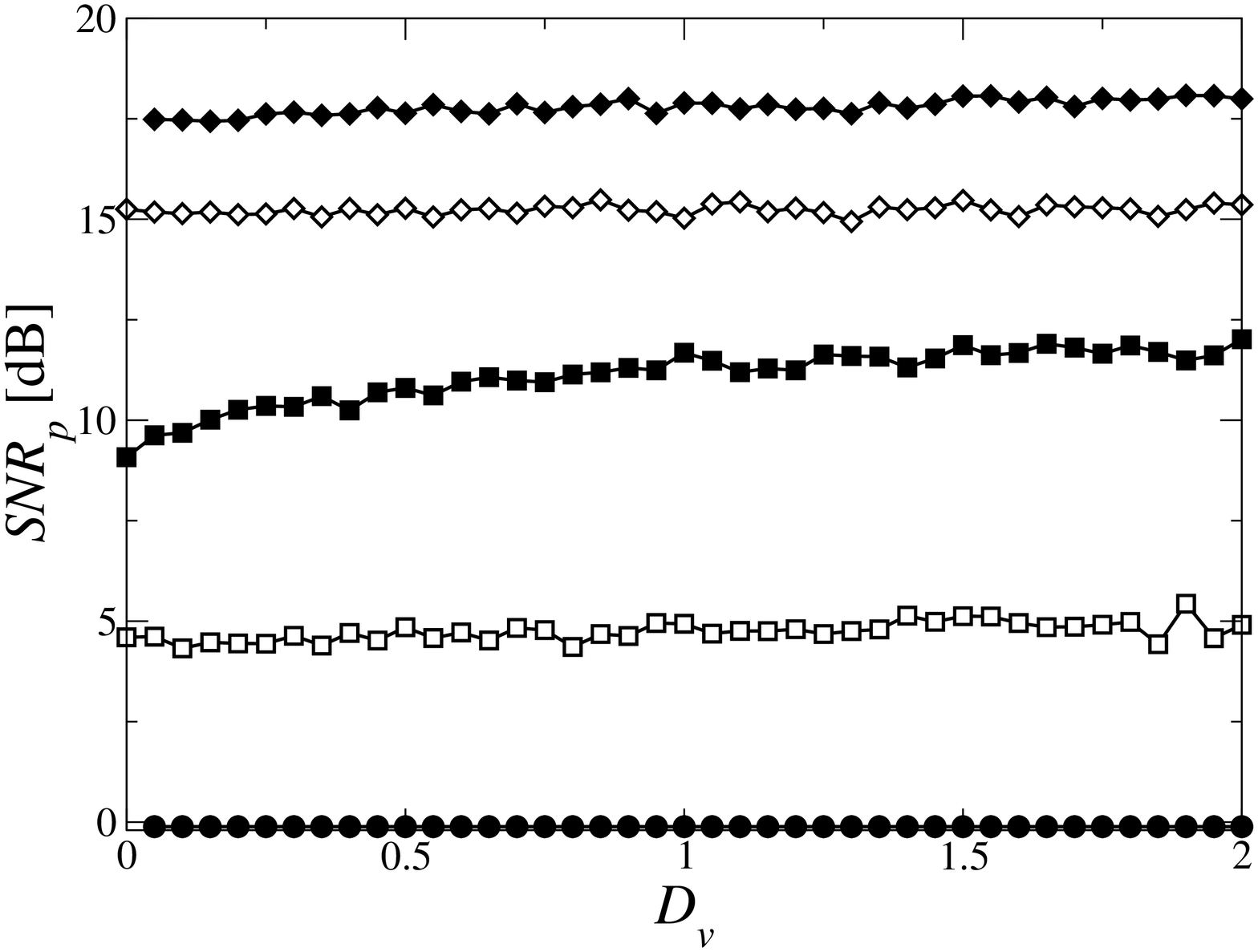}
\caption{SNR vs. $D_v$, the diffusiveness in the inhibitor variable
$v$. The parameters are $\delta\phi= 0.4$, $D_u= 0.3$, $\omega =
2\,\pi / 3.2$, $N=51$.  $\gamma = 0.1~(\bigcirc), 0.32~(\square),
1.0~(\diamond)$, while the white symbols represent $h=2$ and the
black ones, $h=0$.}\label{SNR-Dv}
\end{figure}

Next, in figure \ref{SNR-Dv}, we present the results for the SNR as
function of $D_v$, the activator diffusivity, for different values
of $\gamma$, and for $\delta\phi= 0.4$, $D_u= 0.3$, and $N=51$.  We
see that for $h \geq 0$ the response is more or less flat, however,
it is again apparent the SNR's enhancement for $h>0$. For $h<0$ we
see that the system's response decays very fast with increasing
$D_v$. This effect could be associated to the fact (as found in
those cases where the NEP is known \cite{extend2,extend3}) that in
the underlying NEP the bistability is lost as a consequence of the
disappearance of some of the attractors \cite{I0}.

\begin{figure} 
\centering
\includegraphics[width=4.5cm,angle=-90]{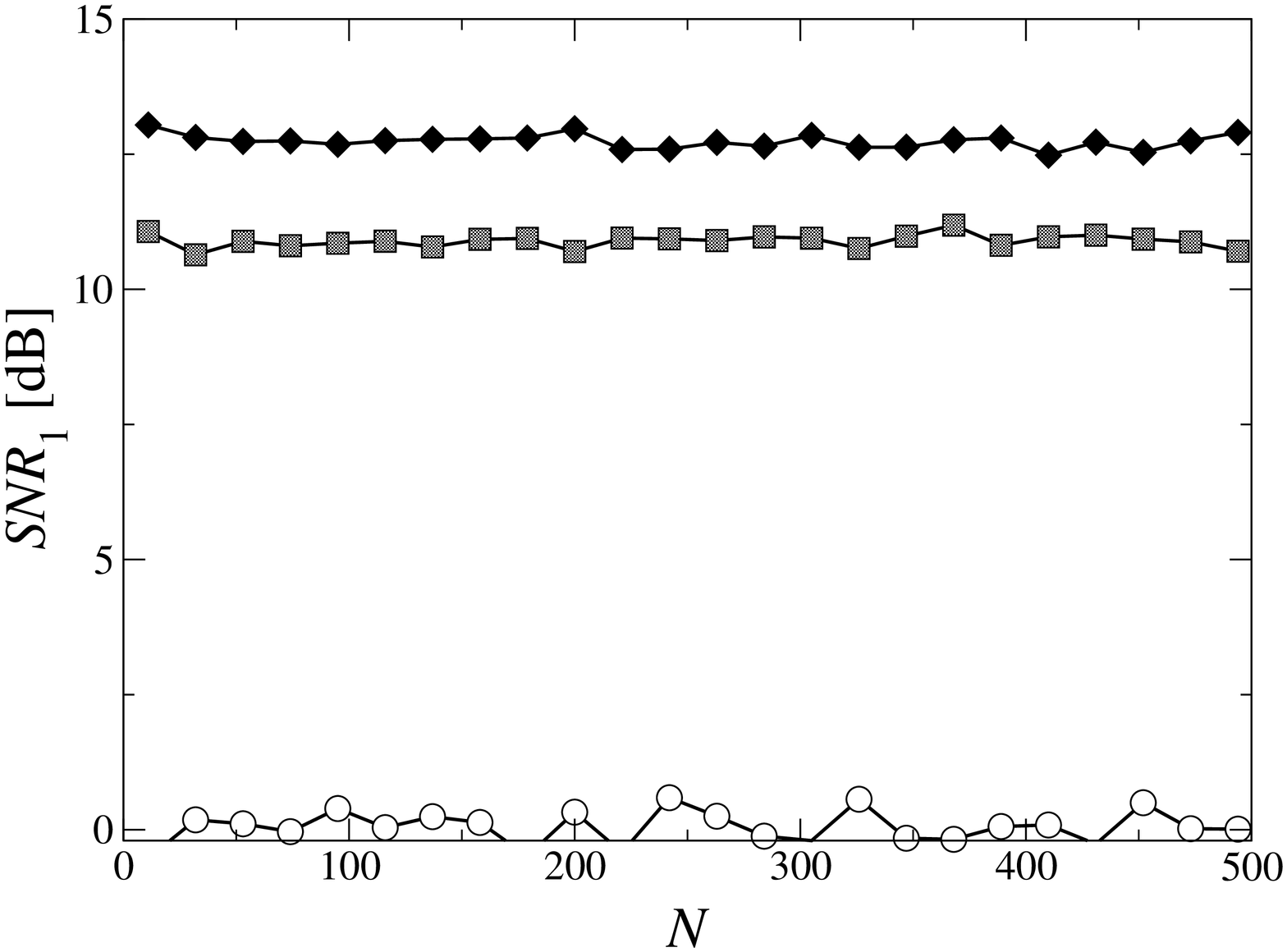}
\includegraphics[width=4.5cm,angle=-90]{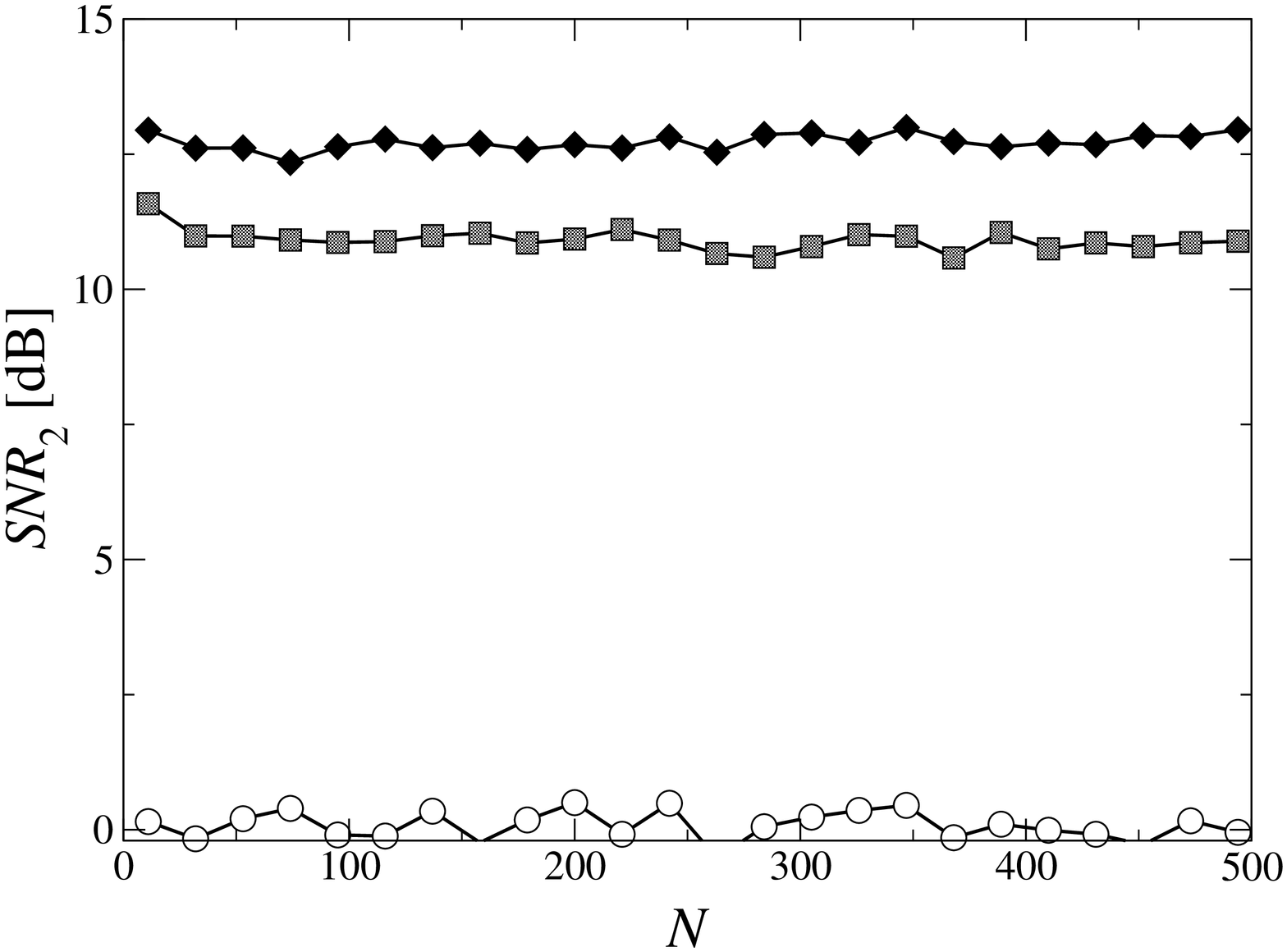}
\includegraphics[width=4.5cm,angle=-90]{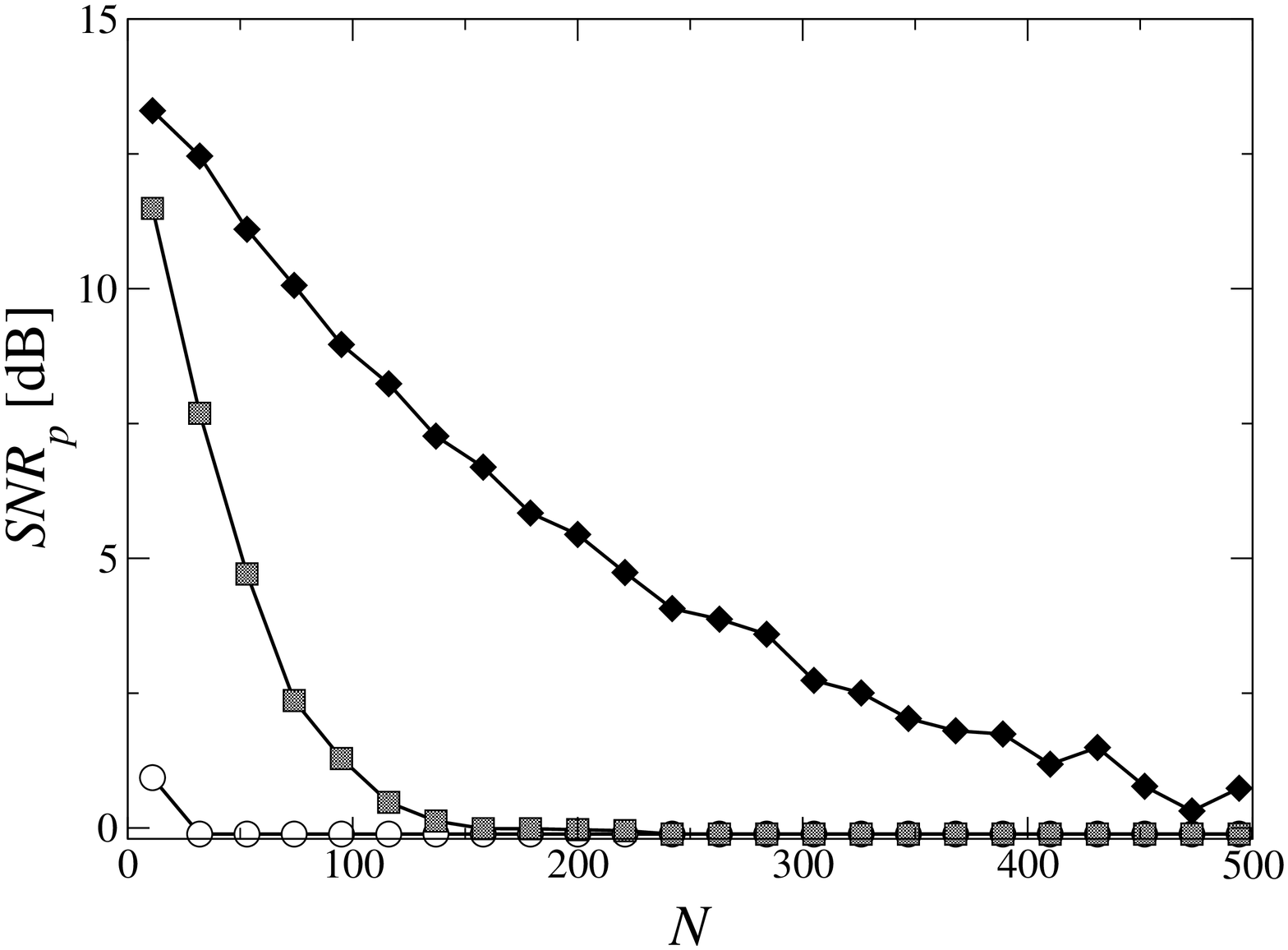}
\caption{SNR vs. $N$, the system size, for the three different
measures we use. The parameters are $\delta\phi= 0.4$, $D_u= 0.3$,
$D_v= 1.$, $\gamma = 0.32$, $h=-2~(\bigcirc), 0~\hbox{(grey
squares)}, 2~(\blacklozenge)$.}\label{SNR-N}
\end{figure}

Finally, in figure \ref{SNR-N} we depict the same three SNR's
measures but as a function of $N$, the system size. For the two
measures $SNR_1$ and $SNR_2$, we see that, for different values of
$h$ and $\gamma$, the response is very flat, and do not seems to be
too much dependent on $N$. It is clear that there is an increase of
the response when $h$ increases. At variance, for $SNR_p$, the
dependence to $N$ is apparent: the SNR decays to zero, in a fast or
slow way, depending of $h=0$ or $h>0$. Here $\delta\phi= 0.4$, $D_u=
0.3$, $D_v= 1.$, $\gamma = 0.01, 0.1, 0.3$.

\section{Conclusions}

We have analyzed a simplified version of the FitzHugh-Nagumo model
\cite{extend3,I0,FHN}, where the activator's diffusion is
density-dependent. Such a system, when both diffusions are constant
(that is: $D_u >0$ and $D_v = 0$), has a known form of the NEP
\cite{extend3}. However, in the general case we have not been able
to find the form of the NEP (but the idea of such a NEP is always
\textit{underlying} our analysis) and we have to resort to an
analysis based on numerical simulations.

Through the numerical approach we have studied the influence of the
different parameters on the system response. From the results it is
apparent the enhancement of the output SNR as $h$, the selectivity
parameter, is increased. This is seen through three different ways
of characterizing the system's response. We can conclude that the
phenomenon of enhancement of the SNR, due to a selectivity in the
coupling, initially found for a scalar system \cite{extend3c} is
robust, and that the indicated nonhomogeneous coupling could clearly
contribute to enhance the SR phenomenon in very general systems.
This phenomenon is also robust to variations of the parameter that
controls the selectiveness of the coupling, up to a point that even
in the case of inhibitory coupling the phenomenon holds.

An aspect worth to be studied in detail is the dependence of the SNR
on $N$, the number of coupled units. In this way we could analyze
the dependence of the so called \textit{system size stochastic
resonance} \cite{haenggi2,SSSR} on the \textit{selective coupling}.
The thorough study of this problem will be the subject of further
work.

\vspace{0.25cm}

{\bf Acknowledgments:} HSW thanks to the European Commission for the
award of a {\it Marie Curie Chair} at the Universidad de Cantabria,
Spain.





\begin{thebibliography}{}


\bibitem{RMP} L. Gammaitoni, P. H\"anggi, P. Jung, F. Marchesoni,
Rev. Mod. Phys. \textbf{70}, 223 (1998).

\bibitem{biol} J.K. Douglas {\em et al.\/}, Nature \textbf{365},
337 (1993);\\ J.J. Collins {\em et al.\/}, Nature \textbf{376}, 236
(1995);\\ S.M. Bezrukov, I. Vodyanoy, Nature \textbf{378}, 362
(1995).

\bibitem{sch} A. Guderian, \textit{et al}; J. Phys. Chem. {\bf 100},
4437 (1996);\\ A. F\"{o}rster, \textit{et al}, J. Phys. Chem. {\bf
100}, 4442 (1996);\\ W. Hohmann, \textit{et al}, J. Phys. Chem. {\bf
100}, 5388 (1996).

\bibitem{extend1} J.F. Lindner {\em et al.\/}, Phys. Rev. E
\textbf{53}, 2081 (1996);\\ J.F. Lindner {\em et al.\/}, Phys. Rev.
Lett. {\bf 75}, 3 (1995).

\bibitem{extend2} H.S. Wio, Phys.\ Rev.\ E \textbf{54}, R3045
(1996);\\ F. Castelpoggi, H.S. Wio, Europhys. Lett. \textbf{38}, 91
(1997);\\ \textit{ibidem}, Phys. Rev. E \textbf{57}, 5112 (1998);\\
S. Bouzat, H.S. Wio, Phys. Rev. E \textbf{59}, 5142 (1999).

\bibitem{extend3} H.S. Wio, S. Bouzat, B. von Haeften, in
\textit{Proc. 21$^{st}$ IUPAP International Conference on
Statistical Physics, STATPHYS21}, A.Robledo, M. Barbosa (Eds.),
Physica A \textbf{306C} 140-156 (2002).

\bibitem{GR} R.Graham, in \textit{Instabilities and Nonequilibrium
Structures}, Eds. E. Tirapegui and D. Villaroel (D. Reidel,
Dordrecht, 1987);\\ R. Graham, T. Tel, Phys. Rev. A \textbf{42}, 4661
(1990);\\ R. Graham, T. Tel, in \textit{Instabilities and
Non-equilibrium Structures III}, E. Tirapegui, W. Zeller, eds.\
(Kluwert, 1991);\\ H.S. Wio, in \textit{4th.\ Granada Seminar in
Computational Physics}, Eds.\ P. Garrido, J. Marro (Springer-Verlag,
Berlin, 1997), pg.135.

\bibitem{I0} G. Iz\'us \textit{et al}, Phys. Rev. E \textbf{52},
129 (1995);\\ G. Iz\'us {\em et al\/}, Int. J. Mod. Physics B
\textbf{10}, 1273 (1996);\\ D.H. Zanette, H.S. Wio, R. Deza, Phys.
Rev. E \textbf{53}, 353 (1996);\\ F. Castelpoggi, H.S. Wio, D.H.
Zanette, Int. J. Mod. Phys. B \textbf{11}, 1717 (1997);\\ G. Drazer,
H.S. Wio, Physica A \textbf{240}, 571 (1997).

\bibitem{extend3c} B. von Haeften, R. Deza, H.S. Wio, Phys.
Rev. Lett. \textbf{84}, 404 (2000).

\bibitem{LGN04} B. Lindner, J. Garc\'{\i}a-Ojalvo, A. Neiman, L.
Schimansky-Geier, Phys. Rep. \textbf{392}, 321 (2004).

\bibitem{FHN} H.S. Wio, \textit{An Introduction to Stochastic
Processes and Nonequilibrium Statistical Physics} (World Scientific,
1994);\\  A.S. Mikhailov, \textit{Foundations  of Synergetics I},
(Springer-Verlag, 1990).

\bibitem{Dayan} P. Dayan and L. F. Abbott, \textit{Theoretical
neuroscience: computational and mathematical modeling of neural
systems, Computational neuroscience} (MIT Press, Cambridge, 2001).

\bibitem{1999_nises} J.~Garc\'{\i}a-Ojalvo and J.M.~Sancho,
\textit{Noise in Spatially Extended Systems} (Springer, New York,
1999).

\bibitem{McNM} B. McNamara, K. Wiesenfeld, Phys. Rev. A
\textbf{39}, 4854 (1989).

\bibitem{jung1} P. Jung and P. H\"anggi, Europhys. Lett. {\bf 8},
505 (1989);

\bibitem{jung2} P. Jung, P. H\"anggi, Phys. Rev. A \textbf{44},
8032 (1991).

\bibitem{haenggi1} I. Goychuk and P. H\"anggi, Phys. Rev. E {\bf
61}, 4272 (2000).

\bibitem{gamma1} L. Gammaitoni, F. Marchesoni, E.
Menichella-Saetta, S. Santuci; Phys. Rev. Lett. \textbf{62}, 39
(1989).

\bibitem{bulsladri} A. Bulsara, A. Zador, Phys. Rev. E \textbf{54},
R2185 (1996).

\bibitem{neiman} L. Schimansky-Geier, J.A. Freund, A.B. Neiman,
B. Shulgin, Int. J. of Bif. \& Chaos \textbf{8}, 869 (1997).

\bibitem{noso} C.J. Tessone, A. Plastino, H.S. Wio, Physica A
\textbf{326}, 37 (2003).

\bibitem{TSM} M. San Miguel, R. Toral, \textit{Stochastic Effects
in Physical Systems}, in \textit{Instabilities and Nonequilibrium
Structures VI}, E. Tirapegui, W. Zeller, Eds. (Kluwer Ac.Press,
1999).

\bibitem{haenggi2} G. Schmid, I. Goychuk, and P. H\"anggi, Europhys.
Lett. {\bf 56}, 22 (2001);\\ P.Jung and J.W. Shuai, Europhys. Lett.
{\bf 56}, 29 (2001).

\bibitem{SSSR} G. Schmid, I. Goychuk, P. H\"anggi, Europhys. Lett.
\textbf{56}, 22 (2001);\\ A. Pikovsky, A. Zaikin, M.A. de la Casa,
Phys. Rev. Lett. \textbf{88}, 050601 (2002);\\ R. Toral, C. Mirasso,
J. Gunton, Europhys. Lett. \textbf{61}, 162 (2003);\\ B. von
Haeften, G.G. Iz\'us and H.S.Wio, Phys. Rev. E \textbf{72}, 021101
(2005).

\end{thebibliography}
\end{document}